# Electrohydrodynamics of dielectric droplet collision with variant wettability surfaces

**Nilamani Sahoo [a], Devranjan Samanta [a,*,1],** and

**Purbarun Dhar [b,*,2]**

[a] Department of Mechanical Engineering, Indian Institute of Technology Ropar,

Rupnagar–140001, Punjab, India

[b] Department of Mechanical Engineering, Indian Institute of Technology Kharagpur,

Kharagpur–721302, West Bengal, India

*Corresponding authors:

[1]E-mail: devranjan.samanta@iitrpr.ac.in

[1]Tel: +91-1881-24-2109

[2]E-mail: purbarun@mech.iitkgp.ac.in

[2]Tel: +91-3222-28-2938

## Abstract

In this article, we report experimental and semi-analytical findings to elucidate the electrohydrodynamics (EHD) of a dielectric liquid droplet impact on superhydrophobic (SH) and hydrophilic surfaces. A wide range of Weber numbers ($We$) and electro-capillary numbers ($Ca_e$) is covered to explore the various regimes of droplet impact EHD. We show that for a fixed $We$~60, droplet rebound on SH surface is suppressed with increase of electric field intensity (increase of $Ca_e$). At high $Ca_e$, instead of the usual uniform radial contraction, the droplets retract faster in orthogonal direction to the electric field and spread along the direction of the electric field. This prevents the accumulation of sufficient kinetic energy to achieve the droplet rebound phenomena. For certain values of $We$ and Ohnesorge number ($Oh$), droplets exhibit somersault-like motion during rebound. Subsequently we propose a semi-analytical model to explain the field induced rebound phenomenon on SH surfaces. Above a critical $Ca_e$~4.0, EHD instability causes fingering pattern via evolution of spire at the rim. Further, the spreading EHD on both hydrophilic and SH surfaces are discussed. On both wettability surfaces and for a fixed We,, the spreading factor shows an increasing trend with increase in $Ca_e$. We have formulated an analytical model based on energy conservation to predict the maximum spreading diameter. The model predictions hold reasonably good agreement with the experimental observations. Finally, a phase map was developed to explain the post impact droplet dynamics on SH surfaces for a wide range of *We* and *Ca$_e$*.

## I. Introduction

Hydrodynamics of droplet collision on solid surfaces is important towards understanding and improvement of various engineering applications like spray cooling of hot surfaces, annealing, quenching, spray coating, printing technology, rapid prototyping using polymer droplets, and pesticide deposition [1, 2]. Comprehension of droplet impact hydrodynamics is also of tremendous importance from the point of view of safety and reliability engineering, such as ice nucleation in aero-turbine blades, strength of arc welded components, etc. Drop impact studies are also relevant to several natural processes like rain drop impact on ocean surfaces leading to upward jet and secondary drops, soil erosion kinetics, or rainwater distribution within dense canopies, or underwater noise during rains [1].

Droplet hydrodynamics in the presence of electric field was initiated through the pioneering works by Sir G. I. Taylor [3, 4]. With the advent of superhydrophobic (SH) surfaces in recent years, the suppression mechanics of droplet rebound has become an important issue, to tackle the problem of pesticide deposition on crop leaves, which mostly tend to be superhydrophobic. Prevention of droplet rebound on SH surfaces has been explored via different techniques, such as the use of additives like surfactants [5] or polymers [6-7]. Another method of supressing droplet rebound is by distorting the symmetry of the droplet shape by application of electric field [8, 9]. Electrohydrodynamics (EHD) based control and actuation has been shown to be effective in achieving better performance of droplet-based lab-on-a-chip devices [10-11], atomization processes [12-13], EHD inkjet printing and electronic cooling [14-15].

The conventional droplet impact studies (without any external force field) focus on the spreading and retraction dynamics at different Weber numbers and Capillary numbers. Temporal dynamics of droplet spreading on stainless steel at various temperatures was studied experimentally [16] in a pioneering research. The role of capillary effects and dynamic contact angle was investigated through experiments and numerical simulation [17, 18] and theoretical modelling [19]. Another simulation study [20] highlighted the role of the boundary layer on droplet spreading and retraction dynamics on SH surfaces. Clanet et al. [21] showed that the maximum spreading of a droplet after impact on SH surfaces depends on the initial droplet diameter and *We*. Yonemoto and Kunugi developed an analytical model of droplet spreading dynamics, and further categorized the phenomenon into the capillary and viscous regimes [22].

Droplet rebound suppression on SH surface has also been studied in recent years to mitigate pesticide wastage and resultant soil contamination. Surfactants were found to promote spreading on SH surfaces and retard droplet rebound [5]. In case of non-Newtonian (Boger fluids) droplet impact dynamics, various factors like normal stress opposing the recoil dynamics, slowing down of receding contact line [6] and critical parameters like impact velocity and fluid elasticity [7] were identified to be the factors behind rebound suppression. Recently, Sahoo et al showed that by application of a transverse magnetic field at specific range of impact Weber numbers, droplet rebound can be inhibited [23]. Yun et al. [8] distorted droplets using an electrospray device to attain non-axisymmetric shapes before impact on SH surfaces, and achieved droplet rebound suppression. During retraction of non-axisymmetric (elliptic) shaped droplets, kinetic energy was noted to alternately switch between two perpendicular axes, resulting in insufficient kinetic energy in the vertical direction available for the rebound [8,23]. This finding contrasts with the spherical drops, where the uniform radial contraction results in rebound off the SH surface.

Although the field of droplet electrohydrodynamics was initiated almost more than fifty years ago [3, 4], droplet impact studies in the presence of electric field are few and far between. Khojastech et al. [24] performed numerical simulations of dielectric liquid droplet impact under the influence of electric field and reported that the droplet always deformed in the direction of the electric field. Ryu and Lee [25] showed experimentally that the maximum spreading diameter of a charged droplet on a dielectric substrate is more than an uncharged drop. They suggested that higher interfacial tension between charged drop and solid or ambient gas compared to the uncharged counterpart causes higher spreading diameter. In another experimental study, charged droplet spreading on a conductive substrate was achieved employing a corona discharge assisted technique [26]. Yurkiv et al. [27] modelled the droplet impact onto polarized and non-polarized dielectric surfaces by varying the applied field strength.

In the present article, we have performed experiments to explore the EHD of dielectric droplets impact upon surfaces of different wettability. Our experimental investigations encapsulate a wide range of governing parameters like *We* and electro-capillary numbers $Ca_e$ to explore the various regimes of drop impact dynamics in presence of electric field. Weber number ($We = \frac{\rho V_o^2 D_o}{\sigma}$) is defined as the ratio of inertial force to surface tension, where $V_o$ is the impact velocity (based on free fall assumption from the drop release height), $D_o$ is the diameter of the droplet at the moment of release from droplet dispenser, $\rho$ is fluid density and $\sigma$ is surface tension of the dielectric fluids. Similarly, electro-capillary number ($Ca_e = \frac{\varepsilon_o \varepsilon_r E_o^2 D_o}{\sigma}$) is the ratio of electrostatic force to surface tension force, where $\varepsilon_o$ is the permittivity of the free space, $\varepsilon_r$ is the dielectric constant of the dielectric fluid, and $E_o$ is the applied electric field strength.

The importance of electric field induced Kelvin polarization force at the interface, interfacial behaviour and fluid dynamics of dielectric droplets after impact was studied in detail. The shape of the droplets during spreading and retraction, and their roles vis-à-vis rebound suppression has been explored. A semi-analytical model is constructed to explain the EHD effect on rebound suppression for a fixed *We*. We also focus on the time dependent experimental spreading behaviour and maximum spreading diameter of a dielectric drop on SH and hydrophilic surfaces. An energy principle based semi-analytical model was formulated to compare the experimentally observed maximum spreading diameter. In addition, we develop a phase map that explains the various regimes of post impact electrohydrodynamics for wide range of *We* and $Ca_e$. The results may find strong implications in design and development of utilitarian aspects of dielectric droplet EHD.

## II. Materials and methodologies

The experimental setup is similar to the one used in our previous report [28]. The setup consisted of a digitally controlled, precision droplet dispenser (Holmarc Opto-Mechantronics Pvt. Ltd., India) unit as shown in Fig.1. The experiments were conducted using a precision glass micro-syringe with a 22-gauge steel needle, and volumetric accuracy of ± 0.1 µl. The electrode assembly is made of aluminium metal strips having dimensions of 8 cm (length) x 2 cm (width) x 0.3 cm (thickness). The electric field (horizontal field, refer fig. 1) was generated between the electrodes, which are connected to the regulated high voltage DC power source (Ionics power solution Pvt. Ltd, India), with output rating of 0-10 kV and 0.1% load variation. The substrate is positioned between the electrodes. The needle was carefully positioned for the droplets to fall on the substrate and at the centre of the electrode gap. The gap between the electrodes was maintained 15 mm throughout the study, taking the magnitude of breakdown strength of air into design upper limit consideration. Accordingly, the maximum limit of $Ca_e$~11 was achieved while conducting the impact experiments. A high speed camera (Photron Fastcam SA4) attached with a G-type AF-S 105 mm macro lens (Nikon) was used to capture images at 1024 x 1024 pixels resolution and 3600 fps.

Experiments were performed at ambient conditions (25 °C) on hydrophilic and SH surfaces. Sterile glass slides, thoroughly cleaned with acetone and DI water and then dried in hot air oven, were used as hydrophilic surfaces. The SH surfaces were created on similar glass substrates using a commercial spray (Ultra Tech International Inc., USA). Stable, $TiO_2$ nanocolloids up to 5 wt. % with DI water as base fluid were used to understand the influence of electric field on impact dynamics of dielectric droplets without significantly changing the surface tension. The test fluid properties like density, viscosity and surface tension and size of the dielectric fluid droplets at 25 °C are given in Table 1. Viscosity and surface tension of test fluids were measured using a rotational rheometer (MCR 102, Anton Paar, Germany) and pendent drop analysis, respectively. Each droplet impact experiment was performed thrice to minimize scientific artefacts. The corresponding Ohnesorge number ($Oh = \frac{\mu}{\sqrt{\rho \sigma D}}$) and $Ca_e$ were also calculated. The Ohnesorge number is defined as the ratio of viscous force to inertial and surface tension forces.

**Table 1:** Properties of the dielectric fluids

| Parameter | TiO$_2$ Colloid (1.25 wt. %) | TiO$_2$ Colloid (2.5 wt. %) | TiO$_2$ Colloid (5.0 wt. %) | DI Water |
|---|---|---|---|---|
| Density (kg/m$^3$) | 1031.0 | 1066.5 | 1136.2 | 997.0 |
| Viscosity (Pa-s) | 0.0028 | 0.007 | 0.00975 | 0.001 |
| Surface tension (N/m) | 0.073 | 0.073 | 0.073 | 0.073 |
| Diameter of the drop (mm) | 2.8 | 2.8 | 2.8 | 2.8 |
| $Oh$ | 0.006 | 0.011 | 0.023 | 0.002 |
| $Ca_e$ | 0-7.5 | 0-8 | 0-10 | 0-7 |

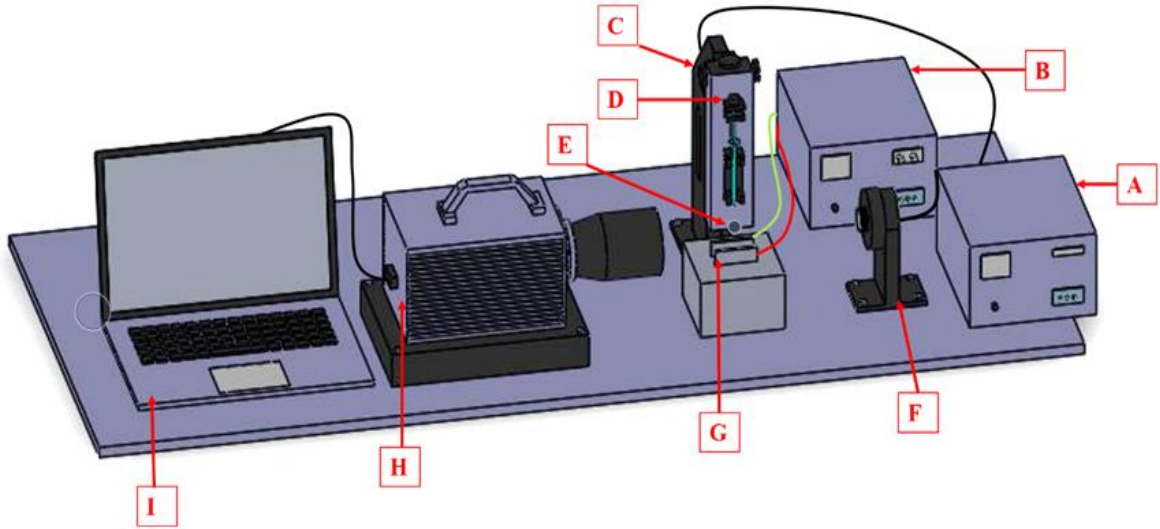

**Fig.1** Schematic of the experimental setup (A) droplet dispenser controller (B) regulated high voltage DC power supply (C) droplet dispenser unit (D) syringe holder (E) pre-impact droplet (F) back light LED array (G) horizontal electrode assembly (H) high speed camera (I) laptop

## III. Results and Discussions

We have discussed different EHD phenomena after impact onto SH and hydrophilic surfaces with varying $Ca_e$ ~0-10 and $We$ ~40-80. Experiments below $We$~40 cannot be conducted beyond $Ca_e$ >0.5 as the needle is present inside the electric field. The external electric field prevents the free fall of droplets from the needle. The various outcomes of the impact dynamics on SH and hydrophilic surfaces are explained as follows:

### A. Droplet rebound suppression on SH surfaces

Although we have covered experiments for Weber number ranging from 40 to 80, $We$~60 was chosen to reveal the various outcomes and the EHD mechanisms responsible for such phenomena. Spreading and retraction dynamics at a fixed $We$~60 and different $Ca_e$ (different electric field strength) have been presented in figure 2. Fig. 2a illustrates the front view of the temporal evolution of droplets after impact on SH surfaces at different $Ca_e$ (columns represent $Ca_e$= (i) 0, (ii) 0.5, (iii) 2, and (iv) 4) for a fixed $We$ ~60. At $Ca_e$=0 (zero field), the droplets exhibit rebound typical to SH surfaces. At low $Ca_e$ (say $Ca_e \leq 0.5$), the droplet retracts uniformly after attaining maximum spreading diameter on SH surfaces (column (i) and (ii) in fig. 2a). For $Ca_e$ ~0 and 0.5, the droplets rebound eventually, although not evident for $Ca_e$ ~0.5 within the selected time frame (figure 2a column ii). With increase in $Ca_e$ ~2, the droplets never leave the surface and rebound suppression is achieved on SH substrates (figure 2a column iii).

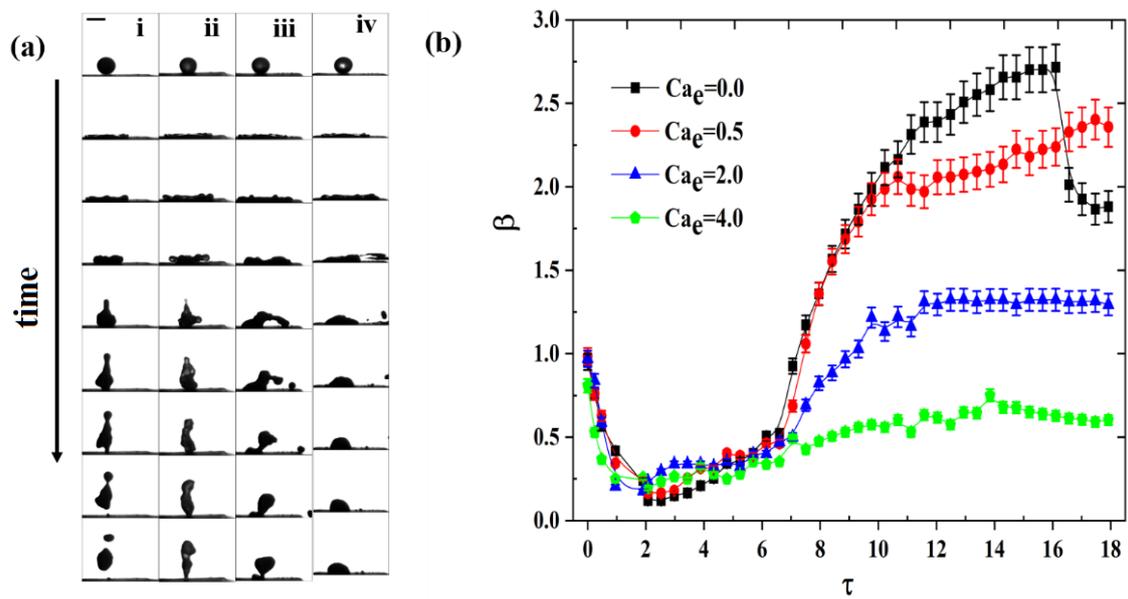

**Fig.2** (a) Front view images of post impact droplet on SH surface. The images are temporally spaced 2.778 ms apart. The scale is 2.8 mm. The electric field acts horizontally across the droplets. Experiments were performed at $We$=60 and different electro-capillary numbers (i) $Ca_e$=0 (ii) $Ca_e$=0.5 (iii) $Ca_e$=2.0, and (iv) $Ca_e$=4.0. (b) Variation of elongation factor with non-dimensional time.

Rebound suppression is more evident for $Ca_e$~4, where the droplet assumes a hemispherical dome shape during retraction and never shows any tendency of lift off from the surface (figure 2a column iii). Figure 3a shows the top view of the droplet EHD for same set of parameters as in figure 2. With increase in $Ca_e$, the droplet retracts faster in the orthogonal direction to the electric field (along the surface but perpendicular to the field lines) and spreads along the direction of the electric field (figure 3a row iii and iv), leading to formation of atypical lead-shaped droplets (fig 3a, $Ca_e$=4, column iv). Instead of radially symmetric droplet retraction, the droplet deforms into thin multiple segments joined by a thin film in between them ($Ca_e$=2 in figure 3a).

Based on the fig. 2a, we have defined an elongation factor (β) as the ratio of the perpendicular height from the point of contact at the solid surface to the top of droplet (L) to the initial droplet diameter ($D_o$). The temporal evolution of the elongation factor is presented in fig. 2b. The non-dimensional time is $= \frac{tV_0}{D_0}$, where t is the elapsed time since the drop touches the substrate, and $V_0$ is the impact velocity. As can be seen from fig. 2b, for $Ca_e = 0$, β increases up to ~2.8 and then sharply decreases near the peak due to a secondary droplet pinch off. In accordance with figure 2a, for increasing $Ca_e$, due to rebound suppression, the elongation factor decreases with increasing $Ca_e$ for a fixed $We$~60 (see fig. 2b). Compared to the zero electric field intensity ($Ca_e$=0); the upper bound of elongation factor is limited to 0.5 for highest value of $Ca_e$~4.

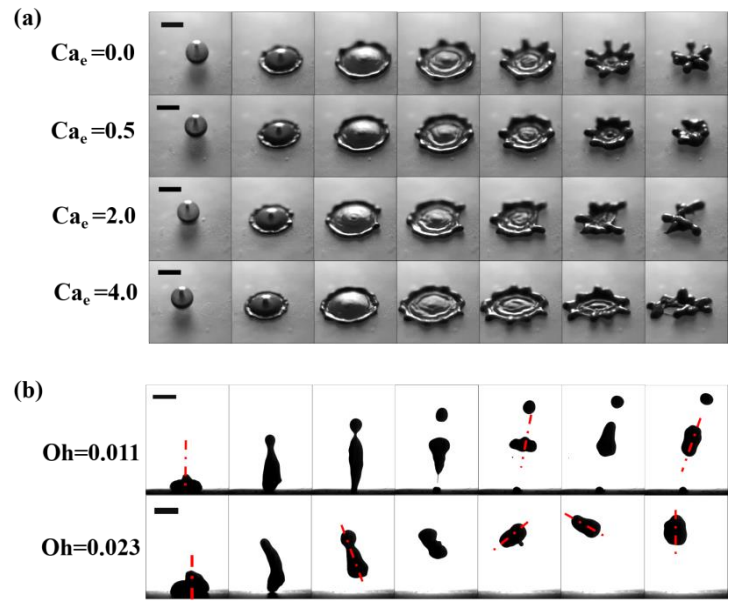

**Fig.3** (a) Top view images of dielectric droplet impacting on SH surface at $We$=60. The time between two consecutive images is 1.385 ms. The scale bar is 2.8mm. The applied electric field is directed horizontally from left to right (b) Rebound morphology of post impact droplet with varying $Oh$ for a fixed $We$=40 and $Ca_e$ =2.0. The time between two consecutive images is 5.54 ms. The scale bar is 2.8mm. At $We$~40, the somersault-like dynamics is more pronounced.

We also observed that the post impact droplet exhibit somersault like behaviour in presence of the horizontal electric field within certain range of $Ca_e$ (0.5<$Ca_e$ ≤4) (fig. 3b). Outside this range, they either bounce off along the vertical direction at low $Ca_e$<0.5, or adhere to the surface for $Ca_e$≥4.0. The rebounding droplet shows angular rotation, while hovering in the confined space between the electrodes at different $Oh$. It also oscillates at different angles with respect to the substrate in presence of the applied electric field, similar to the earlier report by Chen and Bertola [29] about droplet impact behaviour on a superheated surface. The dot-dash red lines (fig. 3b) represent the direction of transient oscillation of bouncing dielectric droplet, and also act as guide to the eye to trace the central axis of the droplet. After bouncing off the surface, the droplet is subjected to an electrical torque caused by the induced dipole due to the directional field. The electric torque provides the physical rotation to the rebounding droplet due to perturbation in the dipole alignment [30].

In order to explain the role of electric field on the retraction dynamics and subsequently the rebound suppression, we have developed an analytical model along the lines of previous reports [31, 32]. During spreading, the droplet attains a thin liquid lamella structure of thickness (h) and radius (R) (measured along the direction of the electric field). Although measurement of radius along the electric field direction is overestimating the actual mean diameter, it can be shown that the electric force term is always dominant over the co-existing surface tension force which contains the maximum spreading factor term. The post impact droplet rapidly retracts forming a rim (with bulbous finger projections) that collects the bulk of the liquid present in the lamella (Fig. 3a). Analogous to an existing study [31], droplet impact dynamics in presence of electric field can be modelled by force balance between the surface tension of the liquid lamella, the electrostatic force, and the inertia of the rim. We can derive the retraction velocity by applying such a force balance principle. The force balance equation is expressed as follows:

$$\frac{d}{dt}\left(m_R \frac{dR}{dt}\right) = F_c + F_k \quad (1)$$

Where $m_R$ is the mass of the liquid in the rim, R is the instantaneous radius of the rim, $F_C$ is the capillary tension acting on the rim, and $F_k$ is the Kelvin polarization force acting on the droplet due to the presence of the electric field.

The capillary force can be expressed following an analytical model proposed by Bartolo et al. [31] as

$$F_c = 2\pi R \sigma_{lv}(1-\cos\theta_r) \quad (2)$$

where $\theta_r$ is the receding contact angle. Prior to impact when the droplet is subjected to the electric field, the electric polarization occurs due to displacement of the paired charges present inside the dielectric fluid. This polarization induces a macroscopic force density, known as Kelvin polarization force density at the interface of the droplet. The Kelvin force density [32] is expressed as follows

$F_k = P \cdot \nabla E_o$  (3)

where P is the polarization density, and $E_o$ is the electric field. Assuming a homogeneous and linear dielectric fluid, the magnitude of polarization density is equal to $(\varepsilon-\varepsilon_o)E_o$. $\varepsilon$ and $\varepsilon_o$ are the permittivity of dielectric fluid of free space, respectively.

$P = (\varepsilon-\varepsilon_o)E_o$  (4)

By substituting the value of P in Eq. (3), we get the force density expression as follows;

$F_k = (\varepsilon-\varepsilon_o)E_o \cdot \nabla E_o$  (5)

Using the vector identity $(\alpha. \nabla \alpha = (\nabla \times \alpha) \times \alpha + \frac{1}{2}\nabla(\alpha.\alpha))$ and since curl of the electrostatic field is zero ($\nabla \times E_o = 0$), the force density can be written as

$F_k = \frac{1}{2}(\varepsilon-\varepsilon_o)\nabla(E_o \cdot E_o)$  (6)

Again, using the vector identity $(\nabla(\phi\beta) = \phi\nabla\beta + \beta\nabla\phi)$, the force density can be further expressed as

$F_k = \frac{1}{2}\nabla[(\varepsilon - \varepsilon_o)E_o \cdot E_o] - \frac{1}{2}E_o \cdot E_o \nabla(\varepsilon - \varepsilon_o)$  (7)

Now, the second term in RHS of the above equation is neglected as the permittivity of the dielectric fluid is uniform and constant. In order to calculate the magnitude of the Kelvin polarization force, Eq. (7) can be integrated over the droplet volume considering a control volume as $F_{kf} = \int \frac{1}{2}\nabla[(\varepsilon - \varepsilon_o)E_o \cdot E_o]dV$, where $F_{kf}$ is the Kelvin polarization force. So, we obtain the magnitude of the force acting on the surface of the droplet as

$F_{kf} = \frac{\pi}{2}\varepsilon_r\varepsilon_o D_o^2 E_o^2$  (8)

where the product of $\varepsilon_r\varepsilon_o \gg \varepsilon_o$ and the surface area considered is that of the initial droplet.

Further, we assume that the inertia of the rim associated with its acceleration is neglected against the capillarity at the onset of retraction phase [19, 31]. Therefore, the inertia of the receding rim can be expressed as:

$F_{rim} = \dot{m}_R \frac{dR}{dt}$  (9)

where $\dot{m}_R = \rho(2\pi Rh)V_{ret}$ and $\frac{dR}{dt} = V_{ret}$, where $V_{ret}$ is the retraction velocity.

Finally, the inertia force can be expressed as:

$F_{rim} = \rho(2\pi Rh)V_{ret}^2$  (10)

Substituting equations (2), (8) and (10) in Eq (1), the final momentum equation can be obtained as

$$2\pi Rh\rho\, V_{ret}^2 = 2\pi R\sigma_{lv}(1-\cos\theta_R) + \frac{\pi}{2}\varepsilon_r\varepsilon_o D_o^2 E_o^2 \quad (11)$$

The moment at which the droplet attains the maximum spreading diameter ($R_{max}$), at which the spreading thickness is $h_{min}$, the final retraction velocity is obtained as follows;

$$V_{ret}^2 = \left(\frac{\sigma_{lv}}{h_{min}\rho}(1-\cos\theta_R) + \frac{1}{2}\frac{\varepsilon_o\varepsilon_r E_o^2 D_o}{\rho h_{min}\psi_{max}}\right) \quad (12)$$

where $\psi_{max}$ is the maximum spreading factor, defined as the ratio of the diameter of the droplet at maximum spread state to the initial pre-impact diameter of the droplet $\left(\frac{D_{max}}{D_0}\right)$.

The receding contact angle ($\theta_r$) and the thickness ($h_{max}$) are considered at the instant of maximum spread. With increase in the $Ca_e$, the term accounting for the electrical force (second component on the R.H.S. of eqn. (12)) becomes dominant over the term accounting for the surface tension component (first term in RHS of eqn. 12) (see fig. 4). When the magnitude of electrical component in equation (12) exceeds the surface tension counterpart, the portion of the rim facing the electrode is stretched due to the electrical stress developed by the Kelvin polarization force. The rim contracts in the direction orthogonal to the external field during the retraction phase (fig. 3a row (iii) and (iv)), by virtue of capillary retraction. The expansion along the field direction and contraction in orthogonal directional to the field results in lower kinetic energy during retraction compared to uniform radial contraction case, ultimately leading to suppression of the droplet rebound (figure 2a column (iii) and (iv)). It also leads to the formation of leaf-shaped droplets during the retraction phase (fig. 3a) in case of high $Ca_e$.

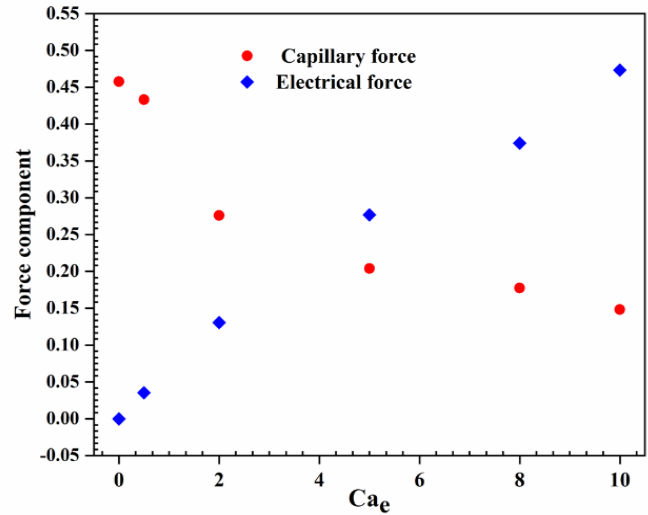

**Fig.4** Influence of the $Ca_e$ on the surface tension and electric force on the droplet. Units are in (N-m)/ kg.

### B. Role of surface wettability on the electro-spreading dynamics

This section discusses the spreading dynamics of the dielectric droplet after impact on both hydrophilic and SH surfaces, at different *We* in presence of the horizontal electric field. The section is categorized as experimental observations on (i) hydrophilic surfaces (ii) superhydrophobic surfaces and (iii) analytical model of the spreading dynamics.

**(i) Hydrophilic surfaces** Fig. 5a and b illustrate the top view images of the dielectric droplet after impact on hydrophilic surface at $We$=60 and 80, respectively. The different rows starting from top to bottom in figs. 5 a-b represent increasing values of $Ca_e$: 0, 0.5, 4 and 8 respectively. For a fixed $We$, with increasing $Ca_e$, the EHD instability sets in to promote the evolution of spires on the rim and stretching of the spread out droplet along the applied field. At high $We$= 80, and $Ca_e$= 8.0, the protrusions or spikes formed at the rim are prominent (fig. 5b, row iv column iii) and are quite reminiscent of the *Rosensweig* or *normal-field instability* in ferrofluids in the presence of magnetic field.

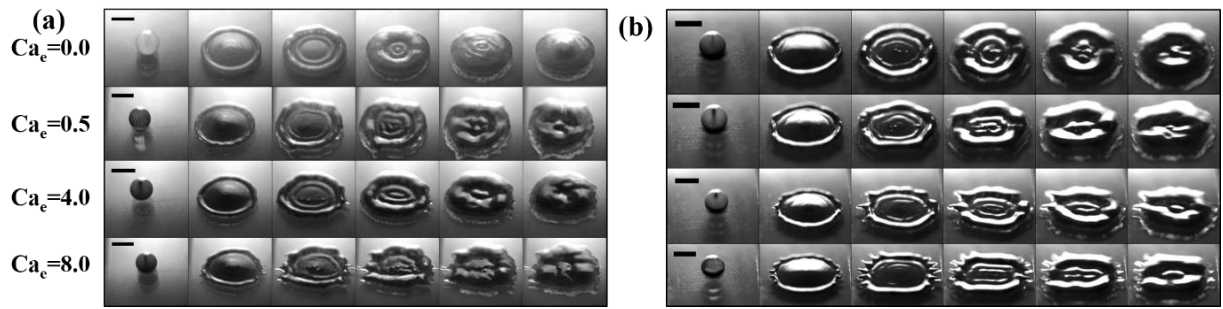

**Fig.5** Top view images of post impact droplet on hydrophilic surface at (a) $We$=60 (b) and We=80. The images are spaced 2.77ms apart. The scale is 2.8 mm. The four different rows starting from top to bottom in each figure (a)-(b) represent values of $Ca_e$ as: 0, 0.5, 4 and 8, respectively.

Subsequently we measured the temporal evolution of spreading factor (figs. 6a and b) based on the top view images of 5a and b. The spreading factor (ψ) is defined as the ratio of the instantaneous diameter of post impact droplet (D) to the initial droplet diameter ($D_o$). During spreading, instead of uniform radial spreading, the droplet elongates more along the direction of the applied electric field compared to the orthogonal direction. With increase in electric field and hence $Ca_e$, the spread out droplet undergoes breakup due to EHD interactions (movies S1 and S2, see supplementary information). Due to the secondary droplet formation and release at τ~9.7 and 14.2, for We~60 and 80, respectively, the spreading factor undergoes sudden decrement at $Ca_e$=8. Beyond $Ca_e$>4, the electrostatic energy strongly enhances the temporal spreading of post impact droplet, thereby showing a steeper slope during the early stage of spreading.

In case of deionized (DI) water, compared to the dielectric droplet, the spire formation around the rim was not observed for any of $Ca_e$ values achieved in our experiments (movie S3). The low surface charge density (due to absence of the dielectric colloidal particles) reduces the Kelvin polarization force acting at the interface. As the magnitude of the polarization force is low compared to the surface tension and viscous forces, the EHD instability is inhibited in case of DI water. It is evident from both fig. 6a and b, that beyond

τ>4, the spreading factor is increasing with $Ca_e$ except for secondary droplet formation and detachment at τ~9.7 and 14.2 for We~60 and 80 respectively and $Ca_e$~8. Overall, from figure 6 it can be inferred that although spreading factor increases monotonously with $Ca_e$ in our experimentally covered range, it hardly shows any difference with respect to Weber number. Thereby, one may propose that in the studied range of $Ca_e$, the inertial effects during both spreading and retraction regimes are dominated by the EHD effects.

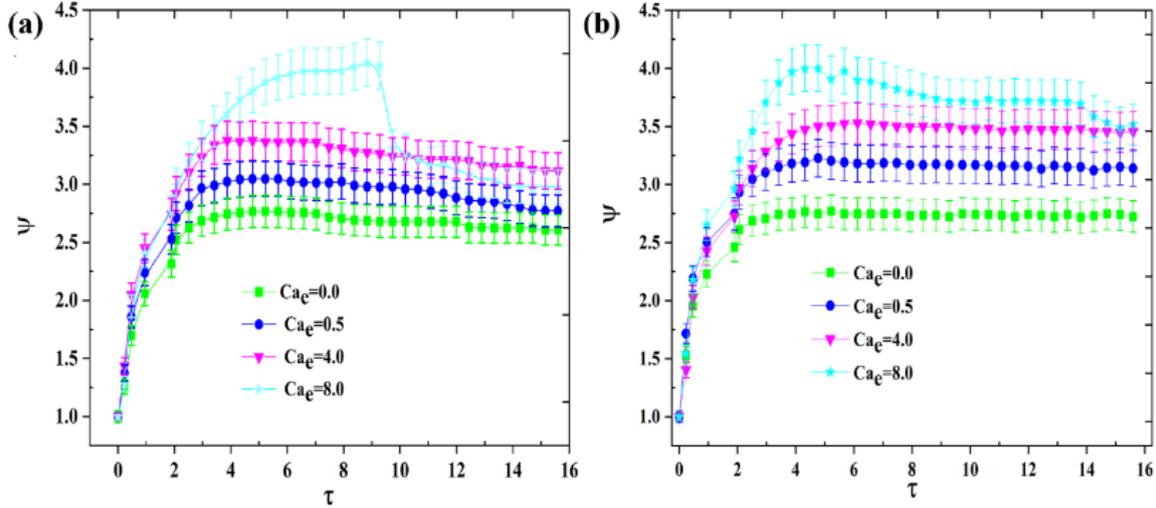

**Fig.6** Temporal evolution of spreading factor on hydrophilic surface at (a) *We*=60 (b) *We*=80.

**(ii) Superhydrophobic (SH) Surfaces**

Figure 7a shows the temporal evolution of droplet spreading on SH surfaces. The retraction of the spread out droplet is delayed with increase in $Ca_e$ (see fig 7b) on a SH surface owing to the generation of electrical stresses by Kelvin polarization force. The spires (manifested through bulbous fingering structures) are formed on the rim due to hydrodynamic instabilities at *We*≥60, similar to the previous studies of drop impact dynamics on SH surfaces in absence of any external force field [33, 34]. The growth of spires is further enhanced in presence of the electric field, which leads to increase in the overall spreading (fig. 7b). At high $Ca_e$~8, due to the enhanced spire growth, portions of droplets reached too close to the electrodes. This leads to the corona discharge across the droplet interface and the electrodes (see the red arrow mark in fifth row, fourth column of Fig.7a). Based on figure 7a, temporal spreading factor (ψ) is presented in figure 7b. In accordance with the increasing elongation of the droplet along the electric field direction for increasing $Ca_e$ in fig. 7a, figure 7b also shows that at τ≥1.0, the spreading factor increases with $Ca_e$.

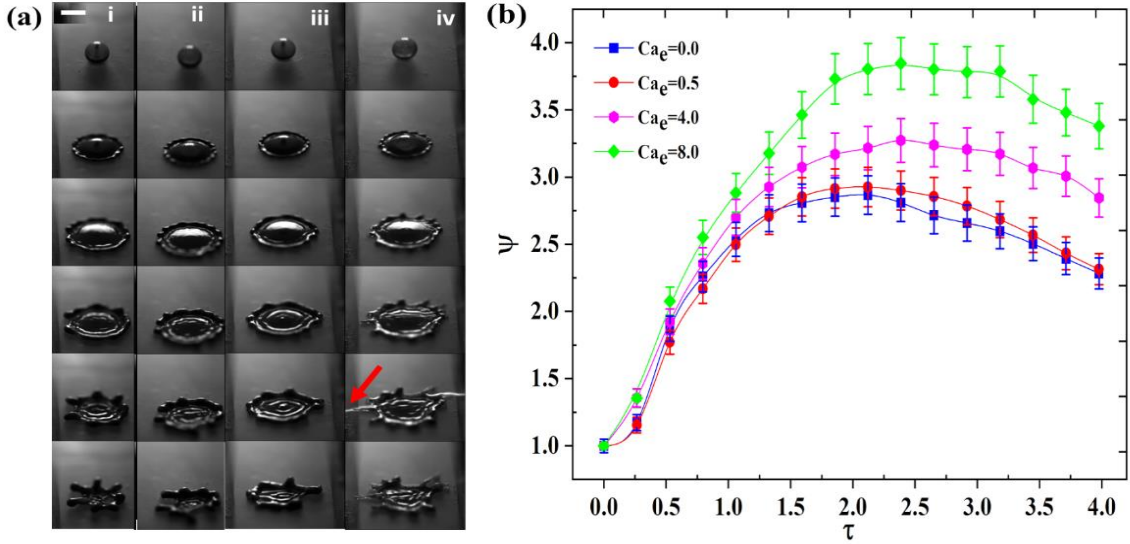

**Fig.7** (a) Top views of post impact droplet and (b) temporal evolution of spreading factor on SH surface at $We=80$. The columns (i)-(iv) represent $Ca_e = 0$, 0.5, 4 and 8, respectively. The images are spaced 1.389 ms apart. The magnitude of scale is 2.8 mm. The arrow is a guide to the eye for corona discharge through the droplet caused by proximity of the bulbous fingering projections to the electrodes.

Subsequently, we studied the role of the Ohnesorge number ($Oh$) on the drop impact EHD. The corresponding changes in physical properties like viscosity, surface tension or density are manifested through the $Oh$. Fig. 8 illustrates the time dependent spreading dynamics of droplet after impact on the SH surfaces for two different Oh values of ~0.011 and 0.023. In case of SH surfaces, for a fixed $We=60$ and $Ca_e=8.0$, the post impact morphology of droplet changes at different Oh. With increase in $Oh$, the viscosity impedes the spreading of droplets post impact at the same $We$ and $Ca_e$, respectively, illustrating the dominance of viscous dissipation over the electrostatic energy.

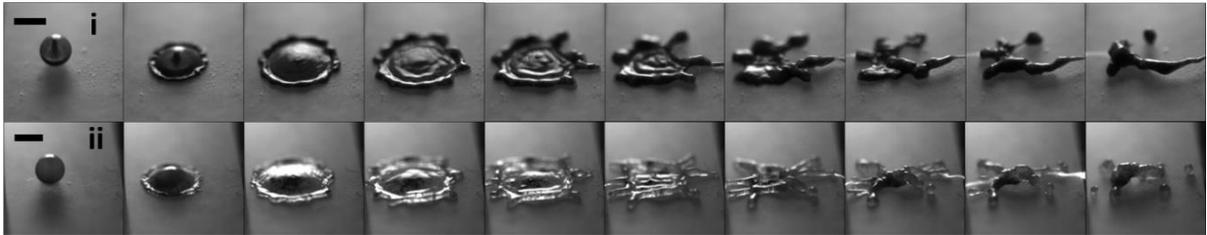

**Fig. 8** Top view images of droplet after impact onto SH surface at $We=60$ and $Ca_e=8.0$ for (iii) $Oh=0.011$ (iv) $Oh=0.023$. The scale is 2.8 mm. The images are spaced at 1.38 ms apart.

### (iii) Analytical model of maximum spreading factor

The present analytical model is based on the model proposed by Yonemoto and Kunugi, hereafter referred as Y-K model [22]. The present approach considers the principle of conservation of energy to deduce the mathematical equation for the maximum spreading factor on SH and hydrophilic surfaces. Prior to impingement, the droplet energy is composed of the kinetic, gravitational, electrostatic, and surface energy components. The gravitational energy component is considered as the drop diameter ~2.8 mm is nearly equal to the capillary length scale for water (~2.7mm). Similarly, the post impact process is associated with interfacial and viscous energies. The wetting phenomenon of variant drop diameter depends upon the motion in the horizontal and vertical directions at the contact line [22]. The present energy equation includes the work done by the adhesive force along the vertical and horizontal direction irrespective the orientation of the external applied field.

The conservation of energy equation pre and post impact is expressed as

$$E_k + E_g + E_s + E_e = E_a + E_v \quad (13)$$

where $E_k$, $E_g$, $E_e$ and $E_s$ are the kinetic, gravitational potential of the droplet, electrostatic and surface energies, respectively. Similarly $E_a$ and $E_v$ are the energy components for adhesion and viscous dissipation after impingement, respectively. Mathematically, each component of the energy equation (Eqn. 13) can be expressed as follows

$$E_k = \frac{1}{2}\rho V_i v_o^2 \quad (14)$$

$$E_g = \frac{1}{2}\rho g h_{min} V_i \quad (15)$$

$$E_e = \frac{1}{2}\varepsilon_r \varepsilon_o E_o^2 V_i \quad (16)$$

$$E_s = \pi D_o^2 \sigma_{lv} \quad (17)$$

$$E_a = \pi R_{max}^2 \sigma_{lv}(1-\cos\theta_d) - \pi R_{max} h_{min} \sigma_{lv} \sin\theta_d \quad (18)$$

$$E_v = \mu \left(\frac{u_r^2}{h_{eff}^2}\right) V_i t_d \quad (19)$$

The parameters described in Eqn. (14-19) are as follows: $\rho$ is the density of the dielectric droplet; $V_i$ is the initial volume of the pre impact droplet; $v_o$ is the impact velocity; $\mu$ is the viscosity of test liquid; $\sigma_{lv}$ is the surface tension of the dielectric liquid; $\theta_d$ is considered as the dynamic contact angle at the moment the droplet attains the maximum spreading diameter; $t_d$ is defined as the ratio of the maximum spreading radius ($R_{max}$) to the initial impact velocity ($V_i$); $h_{min}$ is the droplet height at the moment the droplet attains $R_{max}$. $h_{eff}$ is the effective spreading thickness after droplet impingement with the surface; $u_r$ is the radial velocity at the edge of the liquid film during droplet spreading; $\varepsilon_o$ and $\varepsilon_r$ are the permittivity of free space and relative permittivity of the droplet, respectively; $E_o$ is the applied field strength across the droplet.

The kinetic energy becomes zero after achieving the maximum spreading radius along the solid surface. Along the lines of Y-K model [22], the droplet is assumed to attain a cylindrical shape of diameter $D_0$ and thickness h before spreading on the solid surfaces.

Using conservation of mass principle, the initial radial velocity ($u_{ir,\,mean}$) can be obtained as follows:

$$u_{ir,\,mean} = \frac{3}{8} v_o \quad (20)$$

The present study considers a linearly decaying velocity profile (as explained in literature [22]) to evaluate the maximum radial velocity in the liquid film during the spreading process. In the present approach, the maximum velocity is calculated by assuming the flow between two parallel plates (i.e. the liquid film top surface and the impact wall, similar to shear driven flows). Accordingly, the radial velocity profile [22] is formulated as $u_r = u_{r,\,max}\{\frac{R_{max}-R}{R_{max}-R_o}\}$ and it varies from $u_{r,max}$ at R = $R_o$ to zero at R = $R_{max}$, where $u_{r,\,max}$ is the maximum radial velocity and equal to half of the initial radial mean velocity. Thus, the radial mean velocity can be obtained by integration as follows:

$$u_r = \frac{3}{8} v_o \quad (21)$$

Further, the effective spreading thickness ($h_{eff}$) parameter involved in the viscous dissipation term is approximately evaluated from the experimental study of a wall jet [35] and flow between two parallel plates. It is experimentally reported that the maximum velocity, in case of wall jet, is located near the wall and equal to one fourth of the effective height of the wall jet. Similarly, when the droplet impacts with the solid surface, the effective spreading thickness is expected to be confined between the droplet volume and the wall. It is therefore approximated as the flow between two parallel plates and similar to a wall jet. Hence, based on the above fundamental aspects [22], it is postulated that this thickness can be determined by considering a harmonic mean as follows:

$$h_{eff} = \frac{\frac{h_{min}}{2}\frac{h_{min}}{4}}{\frac{1}{2}\{\frac{h_{min}}{2}+\frac{h_{min}}{4}\}} = \frac{h_{min}}{3} \quad (22);$$

where $h_{min}$ is the thickness of lamella when the droplet attains maximum spreading diameter. Now, the viscous dissipation term can be expressed by substituting the expressions for $u_r$, $t_d$ and $h_{eff}$ as follows:

$$E_v = \mu \left(\frac{u_r^2}{h_{eff}^2}\right) V_i t_d = \mu \frac{81}{64} \left(\frac{R_{max}}{h_{min}^2}\right) v_o V_i \quad (23)$$

Based on the above considerations, the final energy equation is obtained by substituting Eqs. (14-18), and Eq. (23) into Eq. (13) as

$$3(1-\cos\theta_d)\psi_{max}^2 + \frac{81}{64}\frac{D_o^2}{h_{min}^2}Ca_{eo}\psi_{max} - \left(\frac{\rho g r h_{min} D_o}{\sigma} + We + Ca_e + 12\right) = 0 \quad (24),$$

where $Ca_{eo}$ is the capillary number in absence of the external electric field.

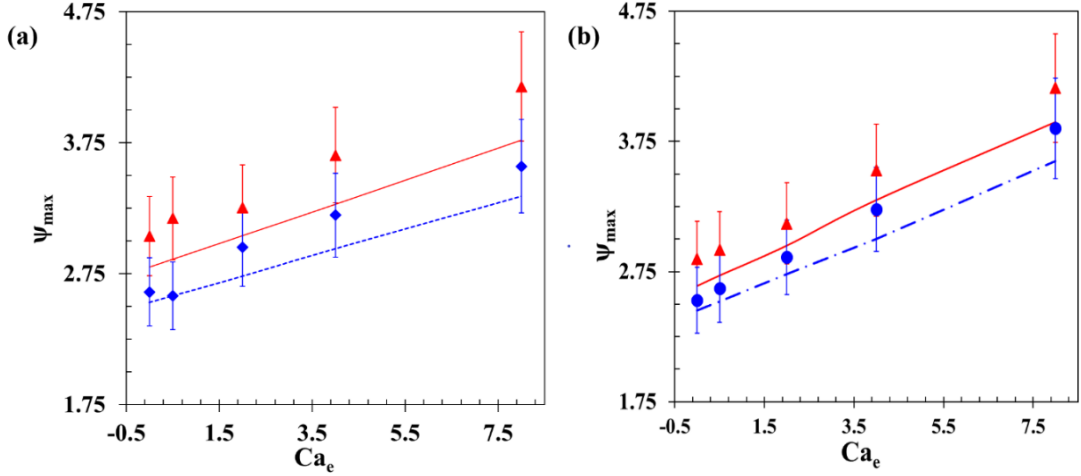

**Fig.9** Comparison between theoretical and experimental maximum spreading factor as a function of $Ca_e$ on (a) hydrophilic surface (b) SH surface. Blue rhombus, red triangle and solid circle are the experimentally observed $\psi_{max}$ values for $We$=40, 80 and 60, respectively. The solid red, blue dotted and long dash-dotted blue lines are the theoretical curves for $We$=80, 40 and 60, respectively.

We have validated our experimental results with the analytical model predictions from Eq. (24). The positive root of the quadratic equation, (Eq. (24)) is only considered in the present study, as it physically signifies the spreading regime. Fig. 9 illustrates the theoretical and experimental values between the $\psi_{max}$ and $Ca_e$ after impact onto hydrophilic and SH surfaces. For dielectric droplets, we observe that the analytical model (Eq. (24)) predicts the experimental $\psi_{max}$ on the hydrophilic surfaces within 10 % error limit. The maximum spreading factor $\psi_{max}$ was increasing with the increment in $Ca_e$ for fixed $We$ on both the surfaces. The under-prediction may be attributed to the numerous assumptions such as uniform flow inside the spreading spat with a linear velocity profile and neglecting the initial charge of the dielectric droplet without considering any residual charge in the theoretical model [17, 22]. In addition, accurate estimation of the viscous dissipation of the droplet in presence of electrical field is difficult due to uncertainty in quantification of the instantaneous contact angles, the localized electroviscous effects, as well as the velocity field within the spreading droplet.

**C. Regime map of droplet impact EHD phenomena**

Fig. 10 represents a regime map of various drop impact EHD observed for the wide range of Weber (20<$We$≤80) and electro-capillary (0.01<$Ca_e$<12) numbers. Experiments were conducted above $We$≥20 in order to avoid the interaction between the needle and electric field. At $We$≤20, the drop impact height is too small and within the domain of influence of the electric field. The field interrupts the free fall of the droplets from the needle in this regime. The impact outcomes noted for different $We$ and $Ca_e$ are as follows:

(1) In R-I region, experiments could not be conducted due to the interaction of the needle with the electric field at *We*<40 and 0.5<*Ca_e*<12.

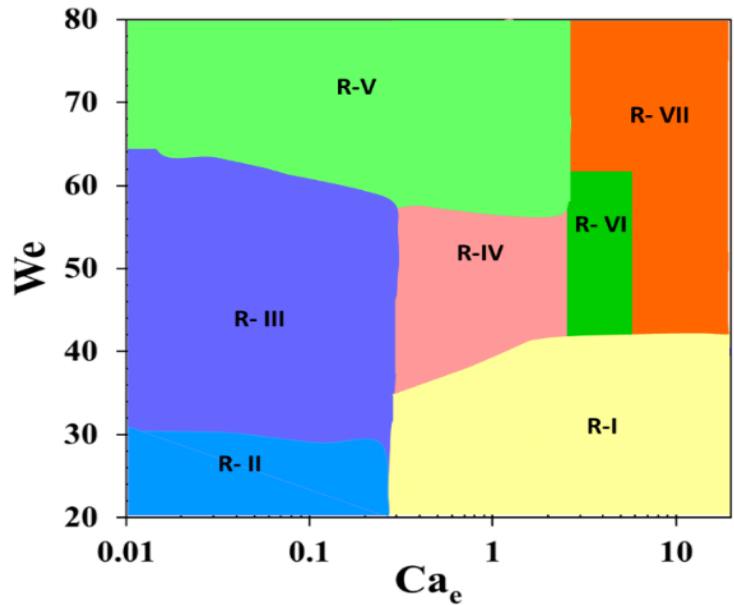

**Fig.10** Regime map of post-impact dielectric drops upon SH surfaces. The coloured regions illustrate the specific impact outcomes at different values of *We* and *Ca_e*.

(2) In the R-II region, for 20<*We*<30 and 0.01<*Ca_e*≤0.5, droplets were observed to rebound. In this regime, the effect of electric field is negligible against the inertia and capillary forces.

(3) The rebound with pinch off is observed in the region R-III at 30<*We*≤65 and 0.01<*Ca_e* <0.5. In this regime, the inertial force is dominant over surface tension and Kelvin polarization forces irrespective of viscous dissipation. After rebounding from the substrate, the droplet always oscillates in the vertical direction normal to the substrate

(4) Regime IV is for 40≤*We*<60 and 0.3<*Ca_e*<4.0. The rebound droplet shows the somersault like behaviour due to generation of the electric torque in presence of the electric field, while hovering in between the electrodes. The bouncing droplet stretches along the dot-dash line as shown in fig. 3b. In this regime, there is a balance between inertial and electrical forces, irrespective of the viscous dissipation.

(5) Again, in the region (R-V) the inertial force is dominant over the electrical force irrespective of viscous dissipation below *Ca_e*<4.0. Therefore, we can observe the crown structure after impact on SH surfaces.

(6) Regime VI shows the rebound suppression behaviour at 4.0≤*Ca_e* ≤6.0. The Kelvin polarization force is dominant over surface tension force and viscous dissipation. Therefore, the spreading droplet exhibits more stretching in the direction of the field, while in the orthogonal direction it spreads relatively lesser. This non-uniform distribution of kinetic energy available during retraction thereby suppresses the rebound.

(7) Beyond $Ca_e$ >6.0 and $We$>60, elongated spires are evolved at the rim due to the EHD instability, leading to corona discharge between the rim and electrodes (the region, R-VII). Though we did not conduct any experiments beyond $Ca_e$>8, it is expected that the EHD instability will play a major role to initiate corona discharge for any $We$>40. Thereby, regimes beyond this may not be important for utilities.

## IV. Conclusions

We have studied the EHD of dielectric droplets upon impact on wetting and non-wetting solid surfaces in presence of the horizontal DC electric field. Phenomena like rebound suppression at high $Ca_e$ for a fixed $We$ was observed while impacting on a SH surface. The spreading droplet undergoes longitudinal stretching due to generation of electric stress by Kelvin polarization force in the direction of the applied field, thereby promoting rebound suppression. For certain $We$ and $Oh$, the droplets rebound and exhibit a somersault-like motion due to electrical torque by perturbation of the dipole moment. We have also developed an analytical model to explain the onset of the rebound phenomena in presence of electric field.

We note that the time dependent spreading diameter increases with the applied field strength after impact on the hydrophilic and SH surfaces. For a fixed $We$, the formation of spires is more pronounced with increasing $Ca_e$. The elongated spire due to hydrodynamic interaction and electrical stress leads to corona discharge across the droplets during spreading upon SH surface. In case of hydrophilic surface, the evolution of multiple spires due to EHD instability occurs, but without corona discharge between the electrodes. Based on the energy balance principle, we have proposed an analytical model for the maximum spreading behaviour. The model agrees reasonably well with the experimental maximum spreading factor. Finally, we develop a phase map encompassing the EHD phenomena on SH surface for wide range of $We$ and $Ca_e$. We believe this study will find strong implications in applications like EHD inkjet printing, electrospraying and electro-spinning, electrokinetics based droplet microfluidic and lab-on-a-chip devices, electro-coating, and so forth.

## Supporting information
The supporting information contains details of three supplementary videos of droplet EHD.

## Conflicts of interest
The authors declare no conflicts of interest with respect to this research work.

## Acknowledgements
NS would like to thank the Ministry of Human Resource Development, Govt. of India, for the doctoral scholarships. DS and PD would like to thank IIT Ropar for funding the present work (vide grants 9-246/2016/IITRPR/144 & IITRPR/Research/193 respectively). Partial funding from the research grant (coded SFI) by IIT Kharagpur is also acknowledged. We also thank